\documentclass[aps,twocolumn,preprintnumbers,amsmath,amssymb,floatfix]{revtex4}

\usepackage{graphicx}
\usepackage{bm}
\usepackage[german,american]{babel}

\newcommand{\figref}[1]{Fig.\ \ref{#1}}

\newcommand{\tabref}[1]{Tab.\ \ref{#1}}
\newcommand{\tls}{TLS}
\newcommand{\dwp}{DWP}
\newcommand{\stm}{STM}

\newcommand{\tg}{\ensuremath{\text{T}_{\text{c}}}}
\newcommand{\md}{MD}

\newcommand{\bmlj}{BMLJ}

\begin{document}


\title{Local Properties of the Potential Energy Landscape of a Model Glass: Understanding
       the Low Temperature Anomalies}

\author{J. Reinisch}
\author{A. Heuer}

\affiliation{Westf\"{a}lische Wilhelms-Universit\"{a}t M\"{u}nster, Institut f\"{u}r Physikalische Chemie\\
and International Graduate School of Chemistry\\
Corrensstr. 30, 48149 M\"{u}nster, Germany}


\begin{abstract}
Though the existence of two-level systems (TLS) is widely accepted
to explain low temperature anomalies in the sound absorption, heat
capacity, thermal conductivity and other quantities, an exact
description of their microscopic nature is still lacking. We
performed computer simulations for a binary Lennard-Jones system,
using a newly developed algorithm to locate double-well potentials
(DWP) and thus two-level systems on a systematic basis. We show
that the intrinsic limitations of computer simulations like finite
time and finite size problems do not hamper this analysis. We
discuss how the DWP are embedded in the total potential energy
landscape. It turns out that most DWP are connected to the
dynamics of the smaller particles and that these DWP
are rather localized. However, DWP related to the
larger particles are more collective.
\end{abstract}

\maketitle

\section{Introduction \label{introduction}}

It is well known that most kinds of disordered solids show
anomalous behavior at very low temperatures as compared to their
crystalline counterparts. Many of the observed features can be
explained by the Standard Tunneling Model (\stm)
\cite{Phillips:1972,Anderson:1972} and its generalization which
is the Soft-Potential Model \cite{Karpov:1983,Buchenau:1991,
Gil:1993, Parshin:1994}. The basic idea of the \stm\ is the
existence of a broad distribution of Two Level Systems (TLS). The
\tls\ can couple to strain and electric fields and therefore
influence quantities like the heat capacity, thermal conductivity,
sound absorption and dielectric response, see \cite{Phillips:1987}
for a review. The \stm\ predicts a linear dependence of the heat
capacity on temperature and a quadratic dependence of the thermal
conductivity on temperature. The \stm\ gives a good general
agreement with experimental results down to temperatures around
100 mK.

So far it has not been possible to derive a theory of the glass
transition or of the low-temperature anomalies from first
principles, i.e. from the Hamiltonian of the glassy system.
This means, that, except for the few cases where
possible \tls\ have been successfully identified by means of
computer simulations \cite{Trachenko:1998,Vegge:2001},
the \stm\ and the recent developments, mentioned above,
are almost purely phenomenological.
In particular researchers wanted to explain why
the nearly constant ratio of the density of \tls\ and their
coupling to phonons, if compared for very different glasses, is so
similar \cite{Pohl:2002}. Already one decade ago this has been
interpreted by Yu and Leggett \cite{Yu:1988,Yu:1989} and Coppersmith
\cite{Coppersmith:1991} as an indication that the observed \tls\
are highly collective excitations of many underlying microscopic
\tls, resulting from the interaction among \tls. In the meantime,
however, it has been shown that the interaction is only relevant
in the mK regime \cite{Natelson:1998,Classen:2000,Rosenberg:2003}.
An alternative scenario has been proposed by Lubchenko and Wolynes
\cite{Lubchenko:2001}. They consider the glass as a mosaic of
frustrated domain walls, separating individual cells. In their
model the collective tunneling process finally involves $O(10^2)$
molecules which only move a fraction $d_L/a$ of a nearest-neighbor
distance $((d_L/a)^2 \approx 0.01)$. Unfortunately, the object of
these theories, namely the
\tls, are still somewhat obscure because experimentally it is very
difficult to characterize their microscopic nature.

Formally, a \tls\ corresponds to a pair of local minima, or a
double-well potential (\dwp), on the potential energy landscape
(PEL). The minima need to have an energy difference less than
$k_B T$ and a small distance in configuration space because
otherwise no tunneling would occur.  One may wonder whether
computer simulations might help to elucidate the relevant
properties of \tls\ and thus to prove their existence. Computer
simulations are strongly limited in several directions:
(a) Typically rather small systems have to be used to analyze the
PEL. This may give rise to significant finite size effects; (b)
Due to finite simulation times the cooling process is extremely
fast so that the resulting glassy structure may be vastly
different as compared to the experimental situation. This may
strongly influence the properties of the \tls\ obtained by
computer simulations; (c) Due to possible imperfections of the
search algorithm to identify \tls\ one may possibly miss a
significant fraction of \tls.

In this contribution we will show that these problems are not
relevant for the problem of locating \tls\ and that it is indeed
possible to obtain detailed and unbiased information about the
nature of \tls. A particular challenge is the systematic search of
\tls. Most of the early computer simulations in this field
\cite{Smith:1978,Brawer:1981,Harris:1982,Guttman:1985,Barnett:1985}
did not attempt to systematically find \dwp. In this work we
present computer simulations on a model glass former (binary
Lennard-Jones) to systematically identify the \tls.  A first step
in this direction has been already published a decade ago
\cite{HeuerSilbey:1993a,HeuerSilbey:1996,Urmann:phd}. At that
time, however, it was not possible to exclude that any of the
above-mentioned problems might hamper the analysis. With
improved algorithms and faster computers this has become possible
nowadays. Furthermore the \tls\ are related to the properties of
the glass transition. Qualitatively, one may say that \tls\ probe
the PEL on a very local scale whereas for the understanding of the
glass transition much larger regions of the PEL are relevant. Here
we would like to mention that in recent years computer simulations
succeeded in extracting many important features of the PEL of
supercooled liquids
\cite{Heuer:1997,Sastry:1998,Sciortino:1999,Buechner:1999,Doliwa:phd,Oligschleger:1995,Schober:2002,Vogel:2003, Wales:2003}.

\section{Technical \label{Technical}}

\subsection{\label{computational} Computational details}
As a model glass former we chose a binary mixture Lennard-Jones
system with 80\% A-particles and 20\% B-particles (\bmlj)
\cite{Kob:1995,Kob:1999,Broderix:2000,Doliwa:2003a}. It is
supposed to represent Nickel-Phosphorous (80\% $^{62}$Ni; 20\%
$^{31}$P) \cite{Weber:1985} but with a 20\% higher particle
density, this system was first used by Kob and Anderson.
The used potential is of the type
\begin{equation}
V_{\alpha\beta}=4\cdot\epsilon_{\alpha\beta}[(\sigma_{\alpha\beta}/r)^{12}-(\sigma_{\alpha\beta}/r)^{6}]
  +(a +b\cdot r),
\end{equation}
with $\sigma_{AB}=0.8\sigma_{AA}$, $\sigma_{BB}=0.88\sigma_{AA}$,
$\epsilon_{AB}=1.5\epsilon_{AA}$,
$\epsilon_{BB}=0.5\epsilon_{AA}$, $m_B = 0.5 m_A$. Periodic
boundary conditions were used and the linear function $a + b \cdot
r$ was added  to ensure continuous energies and forces at the
cutoff $r_c=1.8$.  The units of length, mass and energy are
$\sigma_{AA}$, $m_A$, $\epsilon_{AA}$, the time step within these
units was set to $0.01$. The simulation cell was a cube with a
fixed edge length according to the number of particles and an
exact particle density of $D=1.2$. Molecular dynamics (MD) simulations, using the velocity Verlet
algorithm, have been used to generate independent configurations and
as part of the \dwp\ location algorithm.
For the case of Nickel-Phosphorous the energy unit
corresponds to 933.9 K and $\sigma_{AA}$ is 2.2\ {\AA}.
We analyzed $N=65,\ 2\cdot 65,\ 130,\ 2\cdot 130,\
195,\ 260$ particle systems. The $2\cdot 65$ and the $2\cdot 130$
systems denote two non interacting systems sharing one simulation
box.

It is known for the same system that the dynamics above the
mode-coupling temperature $T_c$ as well as structural properties
like the pair correlation function are basically independent of
system size for $N \ge 65$ \cite{Buechner:1999,Doliwa:2003a}.
Therefore one may hope that even for the very small systems finite
size effects for the \tls\ can be neglected. It will turn out that
this is indeed the case.

\subsection{\label{systematic_locating_of_tls}Systematic Location of \tls}

 We have developed a new algorithm for a systematic search of
\tls. Formally, the problem is to identify two nearby local
minima on the potential energy landscape (PEL) of the system. In
the first step a set of equilibrium configurations is generated
via \md\ simulations at constant temperature $T_{equil}$.
Configurations at different times are taken and used as starting
configurations for subsequent minimization via the Polak-Ribiere
conjugate gradient algorithm, yielding a corresponding set of
local minimum energy structures, denoted {\it inherent}
structures \cite{Stillinger:1982}. In the second step the goal is to locate
nearby minima for these inherent structures. To search for these
nearby minima we proceed as follows: (1) Reset all particle
velocities with random numbers according to a normal distribution
at a fixed temperature $T_{search}$. (2) Perform a fixed number
$N_{search}$ of MD steps in the NVE ensemble. (3) Minimize the
resulting structure. (4) Accept the new minimum if the distance
$d$ and the asymmetry $\Delta$ between both minima in
configuration space (see below for an exact definition) fulfill
$0\approx d_{min}<d< d_{max}$ and $0\approx \Delta_{min}<d \Delta < \Delta_{max}$, respectively. We introduced
very small minimum cutoffs, because minima cannot be distinguished
below numerical precision. For the choice of the
cutoff-values, see below. (5) Repeat this procedure M times for
different initial random numbers. The values of $T_{search}$ and
$N_{search}$ are chosen to give the largest number of nearby
minima per starting inherent structure.

In the final step the transition states for the generated pairs of
minima are located by a modified version of the non-local ridge
method \cite{Doliwa:2003a}. This algorithm is very robust to
identify first-order saddles between pairs of minima. In case that
no such saddle exists the new minimum is dismissed. In particular
this may occur if the path to the second minimum involves
transitions of two (or more) independent \tls\ during the
$N_{search}$ MD-steps. This is the standard situation for very
large systems.
As a result we obtain for every starting minimum a set of minima
characterized by asymmetry $\Delta$, distance $d$ and barrier
height $V$. We explicitly checked that all minima have 0 and all
transitions states 1 negative eigenfrequency.

This algorithm is superior to the \tls-search algorithm, used in
previous works \cite{HeuerSilbey:1993a,Urmann:phd} because no indirect
assumptions are made with respect to the number of particles
participating in a \tls-transition or with respect to the
distance $d$.

Beyond the direct euclidian distance d between configurations (or correspondingly , the distance $d_i$ moved by particle i), we use mass
weighted distances $d_{mw}$ between two
configurations
\begin{equation}
  d_{mw}^2(\vec{r_1}, \vec{r_2})=\sum_i^N (d_{i, x}^2+d_{i, y}^2+d_{i, z}^2)\cdot \frac{m_i}{\bar{m}}.
\end{equation}
Furthermore one can define the mass weighted reaction path approximation
between two minima via
\begin{equation}
  d_{mwrp}=d_{mw}(\vec{r_1},\vec{r}_{trans. state})+d_{mw}(\vec{r}_{trans.
  state},\vec{r_2}).
\end{equation}
Where $\vec{r_1}, \vec{r_2}, \vec{r}_{trans. state} $ are the particle positions of the two minima and the transition state. $d_{rp}$ is defined similarly as $d_{mwrp}$, but without
mass-weighting. The mass-weighted reaction path is introduced because it enters the WKB-term to calculate the tunneling matrix element. For comparison with literature it is necessary to have also direct euclidian distances. In the subsequent analysis we have attempted to employ for every type of analysis the appropriate distance. In any event, due to the similarity of all definitions, none of our results would change on a qualitative level if
different distances had been used.

\subsection{Double-well potentials vs. \tls}

As the ultimate goal one is looking for pairs of minima, i.e.
double-well potentials (\dwp), in the high-dimensional potential
energy landscape with $\Delta < 1K$, corresponding to 0.001 in
LJ-units. They are expected to be relevant for the low-temperature
anomalies around 1K and thus act as \tls. Qualitatively, one would
expect that many more \dwp\ exist with asymmetries of the order of
the glass transition temperature rather than 1K. Restricting the
search to \dwp\ with such a small asymmetry is somewhat problematic
because \tls\ are a very rare species and thus it is very
difficult to find them numerically. Therefore we use as a standard
choice a relatively large asymmetry of $\Delta_{max} = 0.5$. From
the resulting data set (6522 \dwp\ for $N=65$, 2911 for $N=130$ and
428 for $N=260$) we may easily analyze subsets of almost symmetric
\dwp\ which are relevant for the low-temperature properties. As
shown in previous work also the range of very asymmetric \dwp\
contains important information about the properties of \tls. In
particular the choice $\Delta_{max} $ = 20 K (0.02 in LJ-units) is
small enough as compared to the glass transition temperature but
large enough so that we get sufficient statistics. We always
choose $d_{mwrp,max} = 0.8$ for reasons mentioned below. For the
selection of the starting minima we choose $T_{equil} = 0.5$ which
is slightly above the critical mode-coupling temperature of
\tg = 0.45 \cite{Kob:1995,Doliwa:2003a}.

\section{\label{results}Results}

\subsection{\label{r0}Finite size effects and completeness of the search}

\begin{figure}
  \includegraphics[width=8.6cm]{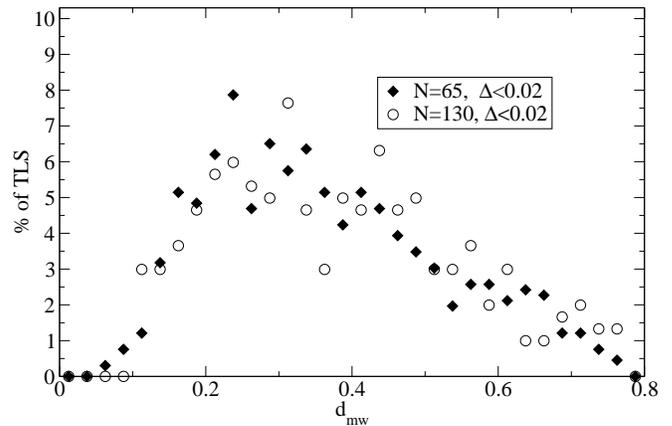}
  \caption{\label{hist20Kvsd} The histogram of the distribution of
  distances for all \dwp\ with $\Delta < 0.02$ (20K for NiP).}
\end{figure}

\begin{figure}
  \includegraphics[width=8.6cm]{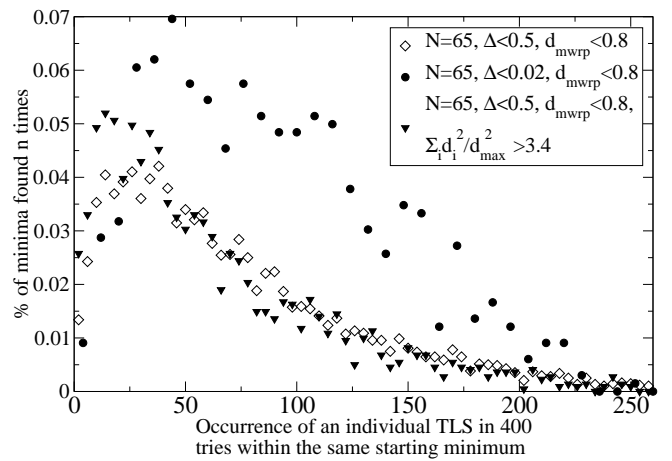}
  \caption{\label{completeness}The histogram of how often a nearby minimum was found for the 65
                particle system after $M=400$ attempts per starting minimum.
        The triangles correspond to data limited to participation
        ratios above average (1/3 of all found).}
\end{figure}

From previous work it is known that the different parameters
$\Delta, d$, and $V$ are strongly correlated. For example one
finds in agreement with intuition that \dwp\ with small $d$
typically have a rather small asymmetry. Thus one would expect
that most \dwp\ which may act as \tls\ at low temperature are
restricted to some region of $d$-values. Therefore we have checked
whether the \dwp\ with $\Delta < 0.02$ indeed display this
restriction in $d$. For $N=65$ we have found 301 \dwp\ in this
range. The result is shown in \figref{hist20Kvsd}.
It turns out that the d-distribution shows a peak around $d=0.3$ and
strongly decreases for larger d. For $N=130$ we get very similar
results. Therefore our
choice $d_{max} = 0.8$ can be justified.
In a next step we elucidate the quality of our search algorithm
and check whether our search of \tls\ is complete within our
specified parameter range. This property is essential to estimate
the absolute number of \tls\ from our simulations.  For this
purpose we have analyzed how often the different minima around a
starting minimum, i.e. how often \dwp, are found during the $M =
400$ attempts. In case that \dwp\ are only found once or twice it is
very likely that many minima are overseen. In contrast, if all \dwp\
are found quite frequently it is unlikely that many \dwp\ are
overseen. This conclusion is based on the assumption that
properties of \dwp\ are continuous, i.e. there is not a second class
of invisible \dwp, strictly separated from the \dwp\ found in our
simulations. For $N=65$ particles the distribution of \tls\ counts
is shown in \figref{completeness} for both $\Delta_{max} = 0.5$
and $\Delta_{max} = 0.02$. Already for the first choice it turns
out that the maximum of the distribution has its maximum at a
count above 15 and the likeliness for lower counts decreases
rapidly. Thus most minima are indeed found. For the \dwp\ with
$\Delta < 0.02$ this effect is even more pronounced. This
comparison also shows that \dwp\ with smaller asymmetry are found
more easily than \dwp\
 with larger asymmetry.

Furthermore we analyzed whether for the system size of $N=65$
finite size effects are present. As already mentioned above,
finite size effects are absent for the dynamics above $T_c$. It
turns out that all properties of \dwp\ (partly presented in this
work, partly in a subsequent publication) ware identical in the
observed N-range between 65 and 260 within statistical noise.
There is, however, one exception. Within the parameter range,
specified above, we found on average 0.65 \dwp\ per starting minimum
for $N=65$ and 0.94 \dwp\ for $N=130$. We used $M=400$ and $M=800$,
respectively. Naively one would have expected twice the number of
\dwp\ for $N=130$. This means that on the per-particle basis the
number of \tls\ is roughly 30\% too small for $N=130$ as compared
to $N=65$. This effect is even more pronounced for $N=260$. Two
reasons are possible: either we have found a significant finite
size effect (which may appear surprising because all properties of
the \tls\ are identical) or this effect may be caused by the
algorithm.

To test the second possibility we performed simulations with two
non interacting systems A and B sharing a box. For determining
minima the relative positions of the two independent systems were
preserved and the energies added. In the combined system the
absolute number of \tls\ is by construction twice as high as in
the elementary system. Interestingly, using the same simulation
parameters as above we also observed a relative decrease of circa
30\% when analyzing the combined system $2 \cdot 65$. Thus it is
very likely that the apparent finite size effect, described above,
is due to the algorithm. This effect can be rationalized.
If a minimum, corresponding to a \tls, is found in configuration
A with a probability of 90\% per attempt and in configuration B
another \tls\ is found with a probability of 10\% per attempt,
they change to 89\% and 1\%, respectively, when both
configurations are considered together and if combined transitions
are ignored (see above). This effect gives rise to a sharp
decrease in probability to detect elementary transitions for
larger systems. Therefore the best procedure is to analyze rather
small systems which, however, are large enough to be void of
relevant finite size effects. In analogy to our results in
\cite{Doliwa:2003a} the system size of $N=65$ seems to be a very good
compromise and, according to \figref{completeness}, allows
one to obtain the absolute number of \tls.

To proceed we counted the number of \dwp\ with $\Delta < 0.02$. It
turns out that one \dwp\ exists per 1000 particles. For even smaller
asymmetry $\Delta < 1$K $(\approx 0.001)$ we observe one \dwp, i.e.
\tls, per 12000 particles. Using the density of NiP one ends up
with $6 \cdot 10^{47} J^{-1} m^{-3}$ \tls. Actually, the parameter
$P$, which can be determined
experimentally and gives an experimental number of \tls,
is roughly smaller by
one decade because the contribution of the individual \tls\ are
weighted by a factor $\Delta_0^2/(\Delta_0^2 + \Delta^2) < 1 $
where $\Delta_0$ is the tunneling matrix element. More
specifically we obtain  $P \approx 1.3 \cdot 10^{46} J^{-1}
m^{-3}$\cite{Reinisch:unpublished} which is close to the value of
$P \approx 1.5 \cdot 10^{46}J^{-1} m^{-3}$, obtained in
\cite{HeuerSilbey:1993a}. Most of the remaining difference with,
e.g. $P = 8 \cdot 10^{44} J^{-1} m^{-3}$ for silicate can be
directly explained by the fact that most glasses have larger
elementary units  like the SiO$_4$ tetrahedra for silicate. This
means that the number of \dwp\ per volume is even smaller, yielding
an estimate very close to the experimental value of silicate.

\subsection{\label{r1}Embedding of \tls\ in the potential energy landscape}
Now we discuss how the \tls\ are embedded into the overall PEL. As
already discussed above, computer simulations suffer from finite
simulation times. As a consequence the starting inherent
structures will have a relatively high energy as compared to
inherent structures the system would have reached if equilibration
at lower temperature $T_{equil}$ close to the calorimetric glass
transition were possible. In order to elucidate this aspect closer
we would briefly like to summarize the previous results for this
LJ-system \cite{Doliwa:2003a}.  (1) Inherent structure energies
range between -304 and -287. The low-energy cutoff, however, is
not due to the bottom of the PEL. Rather it indicates that the
number of minima with even lower energy is so small that they were
not detected during the simulations. (2) More strictly, the
distribution of energies follows the left part of a Gaussian
distribution. This implies that the number of states with higher
energy is always exponentially larger than the number of states
with somewhat lower energy. (3) It can be estimated that the
bottom of the potential energy landscape is around -306. This
number can be estimated from determination of the total
configurational entropy \cite{Doliwa:2003jpc,Sastry:2001}.  Thus
even for very small cooling rates energies would be larger than
-306. (4) The dynamics around and below $T_c$ can be interpreted
as jumps between different traps, denoted metabasins. They contain
a number of individual inherent structures.

\begin{figure}
  \includegraphics[width=8.6cm]{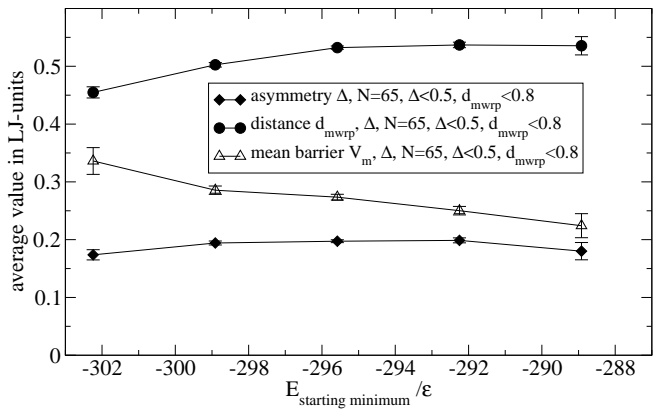}
  \caption{\label{origin_energy_dependence}The dependence of the \dwp\ parameters on the
           energy of the starting minimum.  The lines are a guide to the eye.}
\end{figure}

As our simulations cover a broad range of minimum energies we
checked to which degree the properties of the \dwp\ depend on the
energy of the starting inherent structure.
\figref{origin_energy_dependence} shows the dependence of the
three \dwp\ parameters on this energy. Interestingly, the
dependencies are rather small. In particular there is no
indication to believe that \dwp\ in the relevant low-temperature
energy range between -306 and -302 are very different.
This is in agreement with the observed independence of the
inherent structure dynamics observed by Vogel et al. \cite{Vogel:2003}.
Of course, on a strict quantitative level minor variations
should be taken into account.

\begin{figure}
  \includegraphics[width=8.6cm]{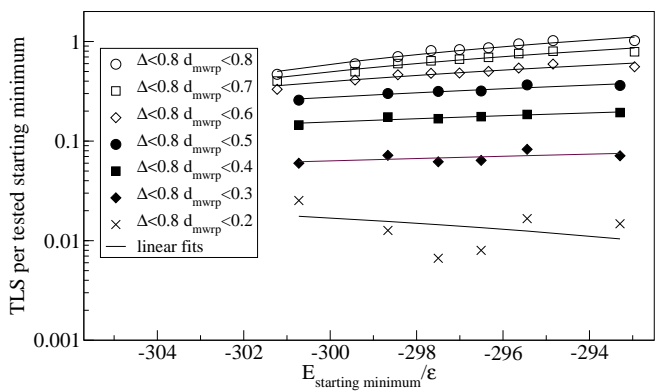}
  \caption{\label{tls_per_origin_vs_E} The dependence of the number of \dwp\ per starting
            minimum on the energy of the starting minimum.}
\end{figure}

\begin{figure}
 \includegraphics[width=8.6cm]{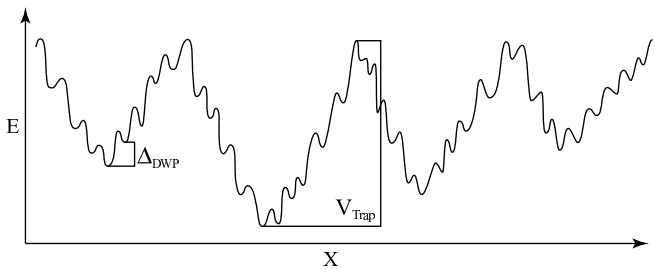}
 \caption{\label{energylandscape}One dimensional sketch of the potential energy landscape.
          The asymmetry $\Delta_{\dwp}$ and the barrier height for the \dwp\ vary
      between 0 and 2 $T_c$. The barrier height for the traps $V_{Trap}$ on the
      other hand varies between 1 and 20 $T_c$.}
\end{figure}

Furthermore we analyzed whether the number of \dwp\ per starting
configuration depends on its potential energy, see
\figref{tls_per_origin_vs_E}.  In order to work with a larger data
set we have included asymmetries up to 0.8. Furthermore we have
analyzed the \dwp\ for different subsets with respect to the
distance $d_{mwrp}$. It turns out that the dependence on energy is
very weak. Even in the most extreme case $(d_{mwpr} < 0.8)$ the
number of \dwp\ changes only by a factor of 2 when comparing
different energies. This small variation nearly vanishes when
restricting ourselves to \dwp\ with small distances. In summary, we
may conclude that on a local scale the properties of the potential
energy landscape do not depend on the height in the landscape.
Evidently this property must break down when taking into account
pairs of minima with larger spatial separation. Because the number
of minima for, e.g., -295 is by a factor of $10^5$ larger than the
number of minima for $-300$ it is evident that for larger $d$ many
more pairs of minima with similar energy can be found for -295 as
compared to -300.

In \figref{energylandscape} we attempt to sketch the PEL of the
LJ-system. The most prominent features are the individual traps.
In contrast, the \dwp\ analyzed in this work correspond to the
little wiggles within the traps. Typical energy scales have been
given in the figure caption showing the separation of transitions
between adjacent inherent structures, already relevant at low
temperatures, and between traps, relevant above the glass
transition.

\subsection{\label{micro} Microscopic nature of \dwp}

\begin{figure}
 \includegraphics[width=8.6cm]{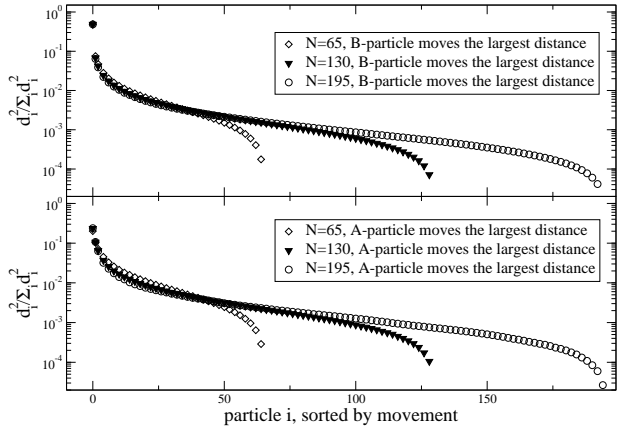}
 \caption{\label{di_distr}The contribution of each particle to the total
           translational motion in a \dwp.  The \dwp\ with an A-particle as central
	   particle (bottom) and a B-particle as central particle (top) show
	   different behavior.}
\end{figure}

Furthermore we have studied the microscopic nature of the \dwp\ in
great detail. As one of the most elementary questions one may ask
how many particle are involved in the translation between the two
minima of a \dwp. To establish an accurate picture of the
microscopic nature of the
  \tls\ we analyzed the number and types of particles participating
  in the transition from one minimum to the other.
  As a matter of fact all particles move when going from one minimum
  to the other. In \figref{di_distr} we show the distribution of
  distances moved by the different particles during the
  transition between both minima of a \dwp.
  The distances are
  sorted according to their actual size. The curvature for the last particles is
  caused by the minimization of the distance of the two minima
  in a finite system. The curves for different systems show
  only small deviations
  for the particles which move a larger distance.
  It can be seen that the first particle, which moves the biggest
  distance, moves much further than the second particle if it is
  a B-particle (small), this is less pronounced if it is an A-particle.
  As the particle which moves the biggest distance can usually clearly
  distinguished from the other particles we denote it
  as {\it central}.
  With this definition we find that in over 90\% of all \dwp\ the
  central particle is a B-particle, although the concentration of
  B-particles is only 20\%. \dwp\ with central B-particles thus dominate
  \dwp\ with central A-particle.

  From these data one may define a participation ratio to estimate
  the number of particles involved in the transition between two
  minima. In case that n particles are moving some fixed distance
  and the other particles do not participate at all one would
  obtain a participation ratio of n. In the present case, of
  course, one has a broad distribution and thus different
  definitions of the participation ratio yield different values,
  as shown in \tabref{tab_part_ratio}. The value for the
  total system varies between 2.1 and 9.5.
  Therefore it is more informative to consider the
  total distribution of translations, as shown in
  \figref{di_distr}.

  \begin{table}
  \begin{center}
  \begin{tabular}{|c|c|c|c|c|}
   \hline
   central                             &
   $1/\langle d_{max}^2/d^2\rangle$    &
   $\langle d^2/d_{max}^2\rangle $     &
   $\langle \sum_i d_i/d_{max}\rangle$ &
   $d^4/\sum_i d_i^4         $          \\
   particle &&&&\\
   \hline
   average&2.14                               & 3.29                                &
   9.50                                & 7.65                                \\
   \hline
   B&2.04                                & 3.01                                &
   8.97                                & 6.95                                \\
   \hline
   A&4.95                               & 6.72                                &
   16.11                                & 16.43                                \\
   \hline
  \end{tabular}
  \end{center}
  \caption{\label{tab_part_ratio}Partition ratios for different definitions and
            for A and B-particles as central particles.
               No mass-weighting has been used for the distances.}
 \end{table}

\begin{figure}
 \includegraphics[width=8.6cm]{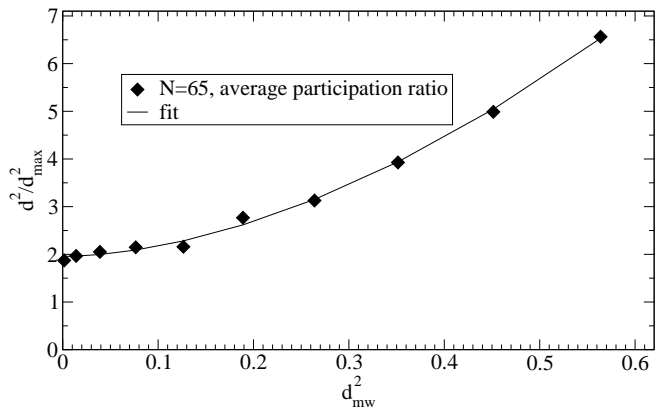}
 \caption{\label{pr_of_d}The participation ratio does significantly dependent
          on the distance of the configurations.}
  \end{figure}

Finally, one may ask the question whether the participation ratio
depends on the properties of the \dwp. Most naturally, one might
think of a relation to the distance between both minima because
the distance directly appears in all definitions of the
participation ratio. In \figref{pr_of_d} the dependence of the
participation ratio (second definition) on the distance is shown.
As expected for \dwp\ with larger distances more particles
contribute to the translation. Actually, in previous work
(\cite{HeuerArtikel} P. 483)  we have
shown for a very similar LJ-system that the slope might decrease
for larger distance, still the participation ratio increases with
distance.

\subsection{Spatial distribution of \dwp}

In the \stm\ it is assumed that the \tls\ are randomly distributed
in space. To check this hypothesis we first calculated the probability
to find two \dwp\ which have the same central particle.
We find that, for both $N=65$ and $N=130$ an enhanced
probability that a single particle is central for two different \dwp.
To
elucidate this aspect further we made a simplified analysis for
the 65 particle system. Instead of considering all pairs of
central particles we considered only B-B-pairs and only those
pairs, which originate from configurations with exactly two
observed \dwp. For independent \dwp\ one would expect that in
$1/13=7.7\%$ of all cases the central particle is identical.
Rather we found (17 $\pm$ 2)\%.  Thus there is indeed a
significantly increased probability that two \dwp\ are spatially
correlated, which can also be interpreted as an increased probability
for triple well potentials.
We find that when omitting \dwp\ with the same central particle the other \dwp\ are
randomly distributed in the system. This was checked by computing
the average distance of the remaining central particles in cases where
more than 1 \dwp\ is present in the system. It turns out that this distance
is within statistical uncertainty identical to the distance of
randomly chosen particles.
The results for the 130 particles system show the same behavior.

\section{Discussion and Summary}
  We have presented a new reliable algorithm to systematically locate \dwp\
  in a model glass former. It turned out
  that the intrinsic limitations of computer simulations do
  not hamper the quality of our results. Thus we can indeed get
  information about \tls, relevant for understanding the
  low-temperature anomalies.

  It turns out that the number of \tls, directly
  obtained from our data, is compatible with the number
  of \tls\ observed experimentally. This conclusion had been
  already drawn from our previous work. This time, however, we
can exclude that possible systematic artifacts hamper our
analysis. Thus it is likely that \tls, responsible for the
low-temperature anomalies, result from elementary two-state
systems rather than from collective excitations.

As a by-product of our simulation we realized that the occurrence
of \dwp\ is not randomly distributed in space. Rather there is a
strong tendency that two \dwp\ are located at the same central particle,
i.e. form a triple well potential. This suggests that some
structural features of the glass favor the formation of \dwp. A
possible candidate is the coordination sphere around the central
particle. Whether or not these
spatial correlations are still significant for \dwp\ with
asymmetries in the Kelvin range would require a much larger set of
\dwp\ than can be found by current simulations.

The properties of the \tls\ reflect local properties of the PEL.
In connection with our previous work on global features of the PEL
it is possible to have a view on the PEL, encompassing the
transport dynamics above and the local dynamics below the glass
transition. It becomes evident, that the \tls\ are part of the
individual traps in which the PEL can be decomposed.

We would like to mention two discrepancies with other work.
One is the magnitude of the absolute movement of a central B or A-particle,
which is found to be quite large:  $d_{B,rp}^2=0.13$ and
$d_{A,rp}^2=0.06$ in LJ units.
These values are about an order of magnitude larger than the estimation given
by Lubchenko and Wolyness \cite{Lubchenko:2001} in their frustrated
domain wall model.
The other discrepancy concerns the participation ratios.
The observed participation ratios, obtained from averaging over all \dwp,
are much
lower than those observed by other groups investigating the PEL,
namely by Oligschleger and Schober in soft sphere glasses and
LJ-systems \cite{Oligschleger:1995,Schober:2002} or Vogel et. al.
\cite{Vogel:2003} in the same system as used in this work. In
\cite{Oligschleger:1995,Schober:2002, Vogel:2003} the authors have
analyzed the transitions between adjacent minima as resulting from
a molecular dynamics trajectory at a given temperature and found
of the order of 20 particles. Comparing this with our average
value of around 3 this seems to be a major difference, as already
stated by Oligschleger and Schober \cite{Oligschleger:1995} in
relation to our earlier work \cite{HeuerSilbey:1993a} where
similar small values have been reported. As shown in
\figref{completeness} this difference is not due to the fact that
our algorithm is not able to identify \dwp\ with large participation
ratios. For a closer discussion of this discrepancy one has to
take care of the actual definition, used to characterize the
participation ratio. In the reported work the two latter
definitions in \tabref{tab_part_ratio} have been used. Thus the
reported values have to be compared to our participation ratios of
about 7-10. There is, however, still a remaining difference of a
factor of 2-3.

This difference can be rationalized by the different methods of
locating inherent structures. In the present work we have
attempted to localize all inherent structures within a certain
distance to the original minimum. This approach was motivated by
the observation that \dwp\ with small asymmetries typically
correspond to nearby minima (see Fig.1) so that \dwp\ with large
distances between the minima are typically irrelevant for the
understanding of the low-temperature anomalies. As a direct
consequence the typical distances between the minima are
significantly smaller found by the present method as compared to
the results reported in
\cite{Oligschleger:1995,Schober:2002,Vogel:2003} which more allude
to the properties of the glass transition. This can be quantified
via the configurational distance $d^C=\sum_i (d_{i, x}^2+d_{i,
y}^2+d_{i, z}^2)^{0.5}$. Comparing our value with the value
reported in \cite{Vogel:2003} we find a difference of a factor of
three.

Following the results, reported in \figref{pr_of_d}, this
difference directly translates into differences of the
participation ratio. Extrapolating the data shown in
\figref{pr_of_d} to distances around unity one ends up with an
increase of the participation ratio by a factor of 2-3 which is
exactly the factor, which was missing above. Thus our present
results are fully consistent with the results discussed in
previous work. We just mention in passing that the situation may
be even more complicated since the \dwp, obtained by Schober et al.
and Vogel et al., have been obtained from MD trajectories. Thus
there exists an implicit weighting by the probability to find
these \dwp. In contrast, in our approach a systematic
determination of all \dwp\ within a specific parameter range has
been conducted. This may hamper the comparison of \dwp, obtained by
the two different methods, even further.

Having in mind the dominance of the small B-particles in the
transition between two minima it becomes obvious that the first
two definitions of the participation ratio better reflect the
dominance of the single-particle character of \dwp\ transitions.
Thus we feel it is more intuitive to speak of 3 rather than of
7-10 particles which dominate the translation between \dwp\
relevant for the low-temperature anomalies. In any event, from a
theoretical point of view it it is the whole distribution in
\figref{di_distr} which fully characterizes the nature of the
translational dynamics between minima in the low-temperature
regime.

One might, of course, ask whether this
result is specific to our binary Lennard-Jones system. Actually,
we already saw that the \dwp\ with central A-particles involve more
cooperative processes. Our preliminary
results for Silica indicate that the participation ratios are
similar to those of the A-particles. In general one might
speculate that for TLS, which mainly contain very similar molecules,
the transition is more collective (like in pure Silica) and behave
like the A-particles
in the present case, while for \tls\
consisting of small  molecules in a matrix the transition
is similar to the more localized process, as seen for the B particles.

Now after validating the methods the next step is to get a
closer insight into the microscopic properties of the \dwp\ and to
perform a direct comparison with experiments on the
low-temperature anomalies. Work along this line will be published
elsewhere.

\begin{acknowledgments}
We like to thank H. Lammert, B. Doliwa, R. K\"uhn, A. Saksaengwijit,
H.R. Schober and M.Vogel for
fruitful discussions and the International Graduate School of
Chemistry for funding.
\end{acknowledgments}
\bibliography{ljtt1_v4}

\end{document}